\documentclass[conference]{IEEEtran}
\IEEEoverridecommandlockouts
\usepackage{amsmath,amsfonts}
\usepackage{array}
\usepackage{textcomp}
\usepackage{stfloats}
\usepackage{url}
\usepackage{verbatim}
\usepackage{graphicx}
\usepackage{cite}
\usepackage{algorithm2e}
\RequirePackage{letltxmacro}
\LetLtxMacro{\LaTeXtextbf}{\textbf}
\LetLtxMacro{\textbf}{\LaTeXtextbf}
\usepackage{cite}
\usepackage{amsmath,amssymb,amsfonts}
\usepackage{graphicx}
\usepackage{textcomp}
\usepackage{caption}
\usepackage{subcaption}
\usepackage[bookmarks=false]{hyperref} 
\usepackage{algpseudocode}
\usepackage{array}
\usepackage{enumitem,lipsum}
\setlist[itemize,enumerate]{leftmargin=*}
\usepackage{etoolbox}
\usepackage{xcolor}
\usepackage{graphicx}
\usepackage{tcolorbox}
\tcbuselibrary{breakable}

\newcommand{\ie}{\textit{i.e. }}
\newcommand{\eg}{\textit{e.g. }}

\ifCLASSINFOpdf

\else

\fi

\begin{document}

\title{LLMChain: Blockchain-based Reputation System for Sharing and Evaluating Large Language Models}

\author{
\IEEEauthorblockN{Mouhamed Amine Bouchiha, Quentin Telnoff, Souhail Bakkali, Ronan Champagnat, Mourad Rabah,\\ Mickaël Coustaty, Yacine Ghamri-Doudane}
\IEEEauthorblockA{ L3i - La Rochelle University, La Rochelle, France \\
{\small\{mouhamed.bouchiha, quentin.telnoff, souhail.bakkali, ronan.champagnat, mourad.rabah, mickael.coustaty, yacine.ghamri\}}@univ-lr.fr}}


\maketitle

\begin{abstract}

Large Language Models (LLMs) have witnessed a rapid growth in emerging challenges and capabilities of language understanding, generation, and reasoning. Despite their remarkable performance in natural language processing-based applications, LLMs are susceptible to undesirable and erratic behaviors, including hallucinations, unreliable reasoning, and the generation of harmful content. These flawed behaviors undermine trust in LLMs and pose significant hurdles to their adoption in real-world applications, such as legal assistance and medical diagnosis, where precision, reliability, and ethical considerations are paramount. These could also lead to user dissatisfaction, which is currently inadequately assessed and captured. Therefore, to effectively and transparently assess users' satisfaction and trust in their interactions with LLMs, we design and develop LLMChain, a decentralized blockchain-based reputation system that combines automatic evaluation with human feedback to assign contextual reputation scores that accurately reflect LLM's behavior. LLMChain helps users and entities identify the most trustworthy LLM for their specific needs and provides LLM developers with valuable information to refine and improve their models. To our knowledge, this is the first time that a blockchain-based distributed framework for sharing and evaluating LLMs has been introduced. Implemented using emerging tools, LLMChain is evaluated across two benchmark datasets, showcasing its effectiveness and scalability in assessing seven different LLMs.
\end{abstract}


\IEEEpeerreviewmaketitle

\begin{tcolorbox}[breakable,boxrule=1pt,colframe=black,colback=white]
\scriptsize Paper accepted at IEEE 48th Annual Computers, Software, and Applications Conference (COMPSAC) IEEE, Osaka, Japan (2024).
\end{tcolorbox}

\section{Introduction}
\IEEEPARstart{L}{arge} Language Models (LLMs) have received a great deal of attention in the last few years due to their surprising capabilities in managing a wide range of Natural Language Processing (NLP) tasks including information retrieval, language understanding, generation, and reasoning~\cite{qinchatgpt,llmjudge}. Despite their impressive capabilities, LLMs such as GPT-3, Llama, and Vicuna~\cite{gpt3,llama,vicuna} exhibit certain challenges that compromise their efficacy. One prominent issue is the manifestation of biases and fairness concerns. LLMs often inherit biases present in their training data, reflecting societal prejudices and stereotypes \cite{trustllm}. Consequently, these models can produce outputs that perpetuate or even exacerbate existing social inequalities. Another limitation arises from the models' difficulty in grasping common sense and contextual understanding. LLMs may struggle to interpret nuances in language, leading to responses that appear nonsensical or detached from real-world knowledge~\cite{llmsurvey}. These behaviors encompass hallucinations, evident in the generation of text that invents or imagines information lacking a factual or coherent basis~\cite{halueval}. LLMs may also display unreliable reasoning~\cite{roscoe}, characterized by a lack of consistent or dependable logical abilities. Furthermore, there is a risk of generating harmful content~\cite{autocorrect}, where LLMs may produce material that is offensive, inappropriate, or potentially harmful. These behaviors can significantly deviate from the expected or desired output, undermining the credibility of LLMs and posing challenges to their widespread adoption. In summary, these flawed actions that diminish trust in LLMs cause users to be cautious about relying on AI-generated content due to its unpredictability and potential for producing incorrect information. They also present hurdles to the utilization of LLMs in critical contexts such as medical diagnosis, legal advice, or sensitive information processing, where accuracy and reliability are essential.

\quad One key way to assess the behavior of LLMs and measure their reliability involves soliciting inputs from users. Individuals can highlight issues they encounter while engaging with AI-generated content~\cite{bridging}. However, this method has two notable drawbacks. First, collecting user feedback is costly as it requires analyzing and categorizing the gathered information. Second, human feedback lacks real-time capabilities as users might not offer immediate responses. This delay hinders prompt evaluation given the absence of instant responses from humans. Therefore, to reduce reliance on human involvement, an alternative strategy consists of employing automatic evaluation methods. These techniques leverage automated feedback~\cite{autocorrect, llmjudge} or language models~\cite{bert, uscore} to evaluate LLMs' performance in a cost-effective way. Despite the efficient processing of language data generated by LLMs, the automatic evaluation metrics they rely on may not perfectly align with human preferences or perceptions, thereby introducing certain limitations. These assessments may fail to capture nuances or qualitative aspects that are crucial for understanding how users perceive the content generated by LLMs~\cite{humeval}. Additionally, existing human and automatic evaluation-based methods face many challenges linked to the lack of transparency and decentralization, as they currently all operate within centralized frameworks. Entities wishing to use LLMs for specific tasks must choose between trusting centralized third-party evaluations or independent testing, which is a costly process that depends on the availability of code and data. Moreover, most of the recent studies concentrate on either human feedback or automated evaluation~\cite{bridging,autocorrect,chiang2023large,gpteval}, missing the opportunity to capture human preferences while enhancing scalability and reducing costs.

\quad To address the above-mentioned issues of evaluating LLMs effectively, dynamically, and transparently, we propose LLMChain, which leverages Blockchain (BC) technology to build a reputation system for LLMs. Blockchains have found extensive use in various trust-related applications such as supply chain~\cite{tcsc}, crowdsourcing~\cite{b9}, and e-commerce platforms \cite{guru}. Its utilization is particularly essential for the development of efficient, decentralized, and transparent reputation systems. These attributes are precisely the qualities we have always envisioned for developing robust reputation systems. Blockchain - known for its resistance to tampering - can be used to track and manage the reputation of various LLMs via smart contracts. LLMChain's primary goal is to help users find the most reliable LLM that meets their specific needs and preferences. Therefore, it allows these individuals to use language models shared by LLM providers and actively participate in the evaluation process.
Additionally, it provides LLM developers with valuable insights, enabling them to enhance and optimize their models by incorporating human feedback. Besides, it is discouraged within reputable organizations for employees to disseminate professional data online or to external entities, a practice that is frequently observed with commercial LLMs. LLMChain aims to address this issue by enabling these organizations to identify open-source LLMs that meet their needs and capabilities for local deployment. This privacy assurance also extends to users who prefer not to share their activities and personal data with third parties. In summary, the contributions of this paper are:

\begin{itemize}
    \item A new reputation-based model. This one is proposed to assess user satisfaction and determine the level of trust associated with each interaction with a language model, via a comprehensive yet scalable evaluation of LLMs' responses (using human feedback and automatic evaluation).
    
    \item A fully decentralized, blockchain-powered platform that enables LLMs to be shared and evaluated thanks to the designed reputation-based model.
    
    \item The preparation of LLMGooAQ\footnote{\href{https://github.com/mohaminemed/LLMGooAQ/}{https://github.com/mohaminemed/LLMGooAQ/}}, a comprehensive dataset encompassing diverse questions and answers across various domains and contexts. This dataset consists of over 100k questions pulled from the large-scale GooAQ dataset and their corresponding answers obtained by performing inference on seven open-source LLMs. 
    
    \item An extensive experimental evaluation with multiple scenarios is performed to demonstrate the effectiveness of the proposed reputation model and the scalability of LLMChain.
\end{itemize}

\section{Related work} \label{sec:relatedWork}

\subsection{LLMs Evaluation} 

\quad To assess the credibility and capabilities of LLMs, several studies have introduced diverse evaluation methods, including pairwise comparison, single-answer grading, or reference-guided grading, employing another LLM as an evaluator. ~\cite{llmjudge, chiang2023large}. These methodologies offer advantages in scalability and interoperability. Nevertheless, it comes with notable limitations: 1) Position Bias, where the evaluator tends to favor the initial model; 2) Verbosity Bias, where the evaluator prefers longer responses over shorter ones; and 3) Self-Enhancement/Promotion Bias, where the judging model prioritizes its own text or that generated from a similar model. Moreover, evaluating a LLM using another LLM appears paradoxical since the evaluator itself is subject to evaluation. On the other hand, alignment-based methods are used to make large-scale alignment research more accessible like OpenAssistant conversations~\cite{openassistant}, which is a corpus of conversations that resemble interactions with assistants, created and annotated by humans. Nonetheless, alignment-based methods face some scalability challenges and annotation expenses. In Core-GPT~\cite{coregpt} and~\cite{assessing}, authors focus on assessing the credibility of LLMs. Core-GPT ~\cite{coregpt} proposes an approach that combines open-access scientific literature with LLMs to improve their reliability and trustworthiness. However, its methodology's scope is limited to two LLMs, ``GPT3.5'' and ``GPT-4'', failing to illuminate the credibility gap between open-source and commercial models. In contrast, the approach proposed in~\cite{assessing} introduces an automated workflow designed to manage an increased number of requests/responses, facilitating the assessment of the credibility of multiple LLMs. In G-Eval~\cite{gpteval}, which is a framework that leverages large language models, used a Chain-of-Thoughts (CoT) and a form-filling paradigm to evaluate the quality of Natural Language Generation (NLG) outputs. G-Eval experimentation involves two generation tasks: text summarization and dialogue generation. However, here again, the methodology is limited to only two LLMs which are ``GPT3.5'' and ``GPT-4''. 

\quad When delineating the prevailing approaches employed to assess the credibility of LLMs, typical challenges become apparent. These approaches lack transparency and decentralization as they all operate within centralized frameworks. To determine the most credible LLM for a specific context, individuals are faced with two alternatives: either relying on centralized evaluations or carrying out tests independently. Additionally, the majority of current studies focus on either human feedback or automated evaluation separately, missing an opportunity to effectively capture human preferences while enhancing scalability and reducing costs.

\subsection{Blockchain-based Reputation Systems} 
\quad The inherent decentralized and tamper-proof nature of blockchain technology provides essential attributes for effective reputation management. Several blockchain-based reputation systems exist, demonstrating the maturity and usability of such solutions for novel applications. TrustChain~\cite{tcsc} is a three-layered blockchain-powered framework used for trust management in IoT-supported Supply Chains. The solution constitutes a service platform operating on a permissioned blockchain network. It leverages smart contracts to automate the computation of reputations and incorporates an incentive mechanism based on rewards and penalties to motivate users toward proper behavior. GuRuChain~\cite{guru}, introduces a blockchain-based service trading platform that incorporates guarantee and reputation at application and consensus layers to foster accountability and trust. It leverages smart contracts to implement the proposed reputation model and manage guarantees using deposits. ValidatorRep~\cite{validatorrep}, is a verification scheme that utilizes blockchain with trust management to foster accountability within crowdsourcing systems. Specifically, this proposal entails a decoupled blockchain model designed for the distinct storage of business transactions and log transactions throughout data interaction. It uses a trust model encompassing the reputation of participants and the trust relationships among them. 
In REPUTABLE~\cite{reputable}, the authors propose a decentralized reputation system for assessing service providers' activity within a blockchain-based ecosystem. The proposed solution integrates a centralized oracle to perform off-chain computations and triggers on-chain smart contracts, impeding the system from achieving complete decentralization. TRUSTD~\cite{trustd} is an ecosystem powered by blockchain and collective signatures, designed to support content creators in garnering community backing for their content. Additionally, it aids users in assessing the credibility and accuracy of these contents.

\quad Therefore, to address the aforementioned challenges related to LLM's evaluation, we believe in the consistency of extending the use of such reputation systems, proposing a novel decentralized framework for evaluating LLMs on open-ended question answering. The proposed concept aims to build a robust and transparent blockchain-based reputation system that merges human evaluation with automated metrics to assess LLMs responses effectively. To our knowledge, this work represents the first study of language model evaluation in a decentralized setting.
\begin{figure*}[t]
   \centering
    \subfloat[System Architecture]{
        \includegraphics[width=9cm, height=6.5cm]{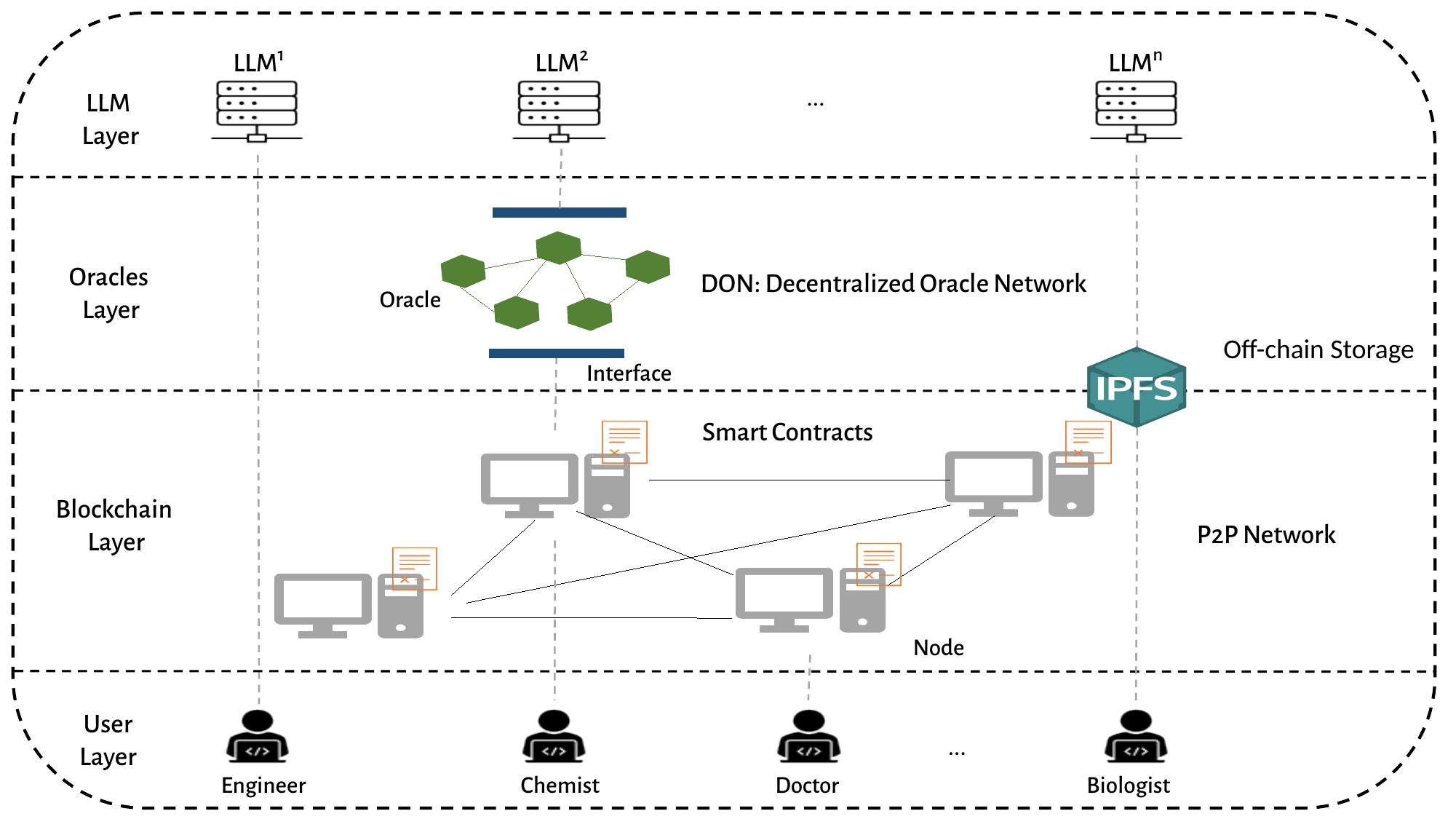}
        \label{fig1a}
    }
    \subfloat[LLM Evaluation Workflow]{
        \includegraphics[width=9cm, height=6.5cm]{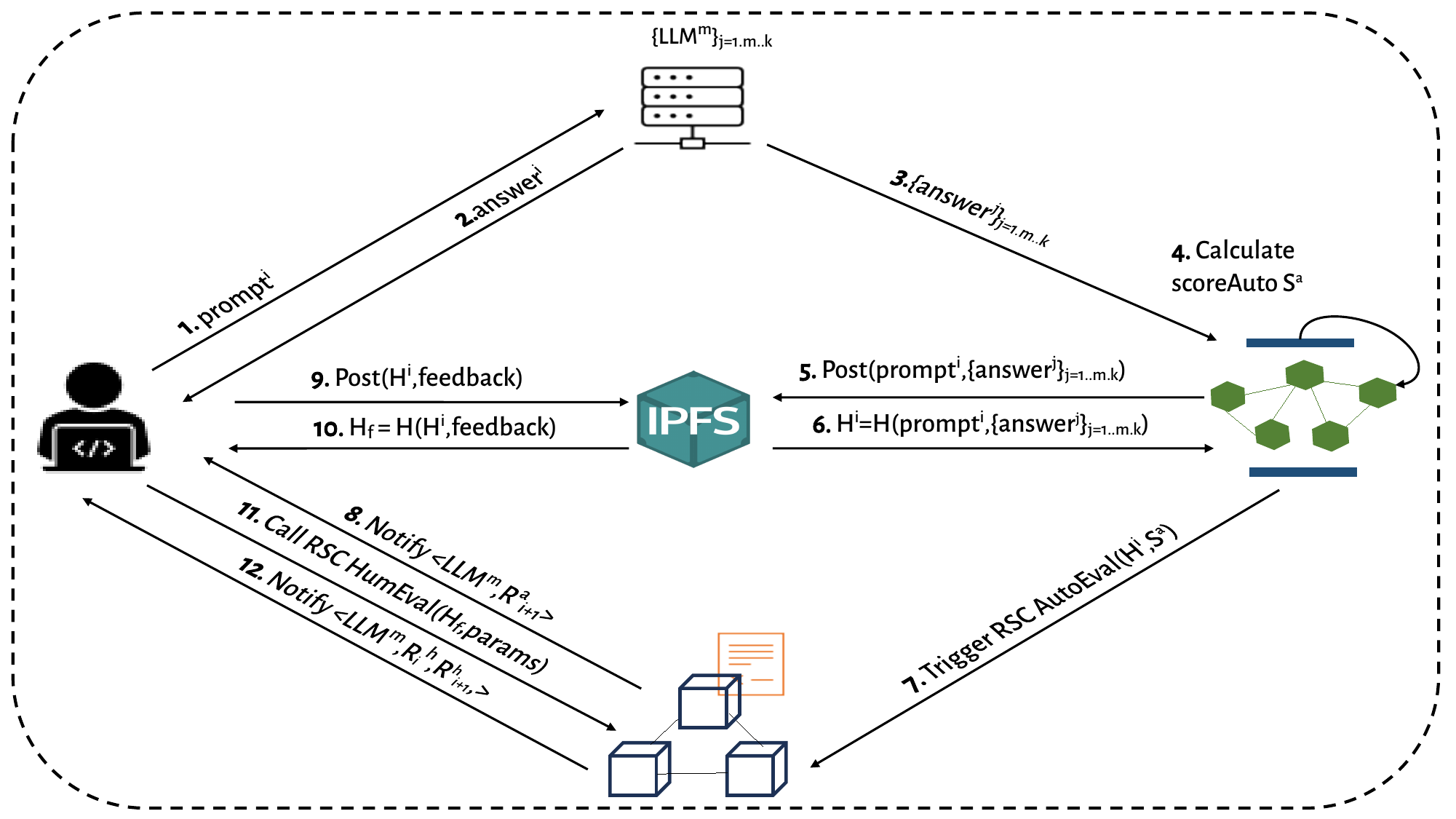}
       \label{fig1b}
    }
\caption{Overview of the LLMChain framework. \ref{fig1a} presents the layered BC-powered architecture. It consists of four main layers: a user layer formed by individuals with different expertise, a BC layer built on a consortium BC managed by LLM providers, and an Oracle layer built up by a decentralized network interconnecting the BC layer with LLMs layer. 
\ref{fig1b} describes the LLMs evaluation process in LLMChain.}
\label{fig1}
\end{figure*}
\section{LLMChain Framework} 
\label{sec:proposedModel} 
\quad In this section, we introduce LLMChain, a Blockchain-powered reputation system for LLM's evaluation. In particular, the proposed framework aims to foster trust in LLMs by amalgamating human feedback and automated evaluations. LLMChain can be seen as a decentralized reputation-based store that allows sharing and evaluating LLMs. It serves a dual role by addressing the needs of users seeking reliable AI assistance, as well as assisting LLMs developers in enhancing the performance and reliability of their models. Fig.\ref{fig1} illustrates an overview of the proposed LLMChain framework.

\subsection{LLMChain Architecture}
\label{subsec:arch}
\quad 
The proposed LLMChain framework is composed of multiple entities distributed over four main layers as depicted in Fig.\ref{fig1a}.

\begin{itemize}
    \item \textbf{User Layer:} is composed of individual participants. Each participant has at least one end-device to interact with the system. Users with different areas of expertise can join the system to use shared, open-access LLMs and provide feedback after engaging with any of the models. This allows users not only to gain insights into the most suitable LLM for their specific domains but also to actively participate in the evaluation process by testing these models and sharing their feedback. 
    
    \item \textbf{Blockchain Layer:} functions as a permissioned blockchain, comprising nodes initiated by LLM providers and/or developers. To participate in the network, an entity must develop and share at least one LLM. LLMChain network employs a consensus mechanism to uphold a uniform ledger copy. We advocate for a reputation-based consensus~\cite{ipot,guru}, leveraging an existing reputation model within the system. Compared to traditional consensus methods, reputation-based consensus offers scalability and enhanced fairness. To further improve the accessibility and performance of our decentralized application, we introduce an InterPlanetary File System (IPFS) \cite{b13} as an off-chain storage system. The core business logic of LLMChain is securely executed via smart contracts deployed over the network and accessed through the submission of transactions. LLM providers benefit from joining the network by gaining full access to LLMChain and, consequently, all the evaluations occurring within the system. This access allows them to accumulate extensive information that will help them to improve and correct their models. 
    
    \item \textbf{Oracle Layer:} comprises Oracle nodes that merge on-chain code with off-chain infrastructure, facilitating the creation of a sophisticated Decentralized Application (DApp). This application responds to real-world events and seamlessly interacts with conventional systems (LLM servers). Hybrid smart contracts deployed across the decentralized Oracle network enable automating the evaluation process. The network intercepts responses from models, conducts off-chain automatic evaluations, and subsequently triggers on-chain smart contracts to update the overall score of the targeted LLM. All of that is achieved in a decentralized and trustless way through the execution of an Oracle protocol~\cite{b31}. 
    
    \item \textbf{LLM Layer:} consists of language models that are administered locally by LLM providers and/or developers. For users who wish to utilize these models for inference tasks, developers need to maintain ongoing access to their shared models. The Oracle network conducts regular checks on the connectivity of these shared models.  Any model that goes offline is automatically removed from the list of running models, keeping the view up-to-date and avoiding interaction with non-operating models.
\end{itemize}

\subsection{LLMs' Evaluation process in LLMChain: An overview} 
\label{subsec:eval}

\quad Unlike centralized frameworks where the evaluation is implemented by a third party, we define end-to-end decentralized evaluation protocols. The proposed protocols are implemented in the LLMChain architecture using smart contracts. The evaluation process consists of three main phases: 


\subsubsection{Registration} To obtain their credentials, including public (address) key and private key, \textit{Users} and \textit{Developers} must register on LLMChain through the Identity Smart Contract (ISC). The registration process can be done in a decentralized, privacy-preserving, and Sybil-resistant way using an IDentity Management Ledger (IDML)\cite{candid}.

\subsubsection{LLM Sharing} LLM developers can add a new model to LLMChain via Reputation Smart Contract (RSC) by calling the $addModel$ function. This creates a new LLM$=\{CID_{llm}, Owner, R_0^{a}, R_0^{h}, R_{0}\}$. $Owner$ is the developer's public key. The initial human $R_0^{h}$, automatic $R_0^{a}$, and weighted reputations $R_{0}$ for the model are calculated as the average values across all existing models in the system. $CID_{llm}$ is the hash of the model's details published on IPFS (\ie The Content Identifier). To ensure the security of LLMChain's smart contract functionalities, we implement role-based access control to manage permissions. This is realized through the Access Control Smart Contract (ACSC). ACSC restricts calling functions by role, for example, it restricts the ability to share models on LLMChain to developers only.
    
\subsubsection{LLM Evaluation} The comprehensive process, spanning from prompt submission to updating the global reputation for the chosen model is illustrated in Fig.\ref{fig1b}. It begins with the user formulating a request intended for a specific $LLM_m$, directly transmitted to the model via a dedicated interface (API). Subsequently, the response from $LLM_m$ is relayed back to the user. To perform \textbf{Automatic evaluation}, the Oracle intercepts both the request and the response. Then, it dispatches identical prompts to other $k$ models $\{LLM_j\}_{j=1,...,k}$, to use their answers as comparative references. Next, it calculates the automatic score for $LLM_m$ using the model described in Sec. \ref{subsec:auteval}. After that, it stores the prompt and its corresponding answers off-chain using IPFS. Finally, it triggers the RSC to update the overall automatic score of $LLM_m$ by calling the $autoEval$ function. Upon receiving the answer, users can opt for direct \textbf{Human evaluation} by calling the $humEval$ function or seek alternative candidate responses to gauge the quality of $LLM_m$'s answer \ie using the shared hash $H^i$, they can retrieve all $k$ answers from IPFS. Once this operation has been completed, the overall weighted reputation score is updated by calling the $updateReputation$ function. Further details on the automatic and human evaluation procedures follow in the next section.  


\section{Reputation Model} \label{sec:reputationFramework}

\quad Human evaluation entails the participation of human experts or users to assess the quality, coherence, and overall adequacy of generated answers. These metrics seek to encompass subjective aspects that automated metrics may overlook \cite{humeval}. Nevertheless, evaluating generated answers through human feedback poses challenges as it relies on users' willingness to offer genuine and immediate feedback. To better address these, we investigate automatic methods, enabling LLMChain to evolve even in the absence of human feedback. In this section, we introduce our reputation model that blends human and automated evaluations. This approach aims to leverage the efficiency and scalability of automated methods while upholding strong alignment through human feedback.

\subsection{Reputation Formulation}
\quad We model the reputation of an LLM as a tuple denoted by $REP=\{R^{a}, R^{h}, R\}$. 
Our approach involves assigning an initial reputation, noted $REP_{0}=\{R^a_{0}, R^h_{0}, R_{0}\}$, to each new LLM. The values of $R^a_{0}$,  $R^h_{0}$, and $R_{0}$ are derived from the average scores of all LLMs in the system.

\quad  The $REP$ tuple undergoes updates after each interaction $i$, following two stages: i) \textbf{Interaction Evaluation}, which involves computing three scores for the targeted LLM - an automatic score $S^a$, a human score $S^h$, and a weighted combination $S^{\theta}$ between both scores - with their respective weights $\omega^a$, $\omega^h$, and $\omega^{\theta}$. ii) \textbf{Global Scores Updating.} each global score $R_{i}$ in $REP$ is updated using a predefined function securely implemented in the RSC contract.  For each $(R_{i}, S_{calc},\omega) \in \{(R^{a},S^a,\omega^a), (R^{h},S^h,\omega^h),(R, S^{\theta},\omega^{\theta})\}$,
 
\begin{equation}  
   \begin{array}{cccc}
\:\:\: \mathcal{U}:& [0,1]\times[0,1]\times[0,1] &\longrightarrow& [0,1]
   \\  &(R_{i},~S_{calc},~\omega) &\longrightarrow& R_{i+1} 
\end{array}
\end{equation}


\subsection{Interaction Evaluation}

\subsubsection{Automatic Evaluation} \label{subsec:auteval} Several studies have demonstrated that embedding-based metrics can effectively match human judgments by considering semantic relevance \cite{bart, disco}. However, their effectiveness is influenced by the quality of the underlying embedding. Consequently, when developing LLMChain, we emphasized a modular framework to retain flexibility in updating the automatic evaluation technique at any time. The metrics we explore to use for the \textbf{Automatic evaluation} requires a minimum of one reference to compute the score $S^a$ (cf. Sec. \ref{subsec:autometrics}). Thus, we propose to use $k$ references, denoted as $\{ref^j\}_{j=1...k}$ to evaluate the answer of the targeted model for better precision. The $k$ references are the answers that the decentralized Oracle gets from the top $k$ models within the context of the prompt. The final score of the answer from the model $LLM$ is computed as follows:

\begin{equation}
        S^{a} = \frac{1}{k} \sum_{j=1}^k  scoreAuto(answer,~ref^j)
\end{equation}

We assess the quality of the automatic evaluation using a weighting function $\omega^a \in [0,1]$. Its outcome varies depending on the average reputation of the models used as references (\ie the better the reputation the higher importance is given). Once this is done, the Oracle triggers the $autoEval$ function in RSC to update the overall automatic score of the $LLM_m$ using the model described in the Sec. \ref{subsec:repupdate}.

\subsubsection{Human Evaluation} \label{subsec:humeval} While it is straightforward to carry out an automated evaluation by measuring the distance/similarity between generated answers, it is less easy to gather information about trust, user satisfaction, completeness, and usefulness of a generated text. Inspired by~\cite{humeval} and~\cite{moritz2018}, our approach involves employing a multi-item scale questionnaire for efficient and scalable human evaluation. Our focus encompasses two types of dimensions (constructs) essential for users to assess text generated by LLM accurately:
\begin{itemize}
\item \textbf{Answer's Constructs:} are the metrics that allow the evaluation of the quality of a single answer/response (\ie calculate $S^h$). To do so, we employ three metrics. First, the \textbf{Reliability}, denoted as $A_t$, evaluates the trustworthiness of the provided answer. Then, the \textbf{Completeness}, denoted as $A_c$, measures the comprehensiveness or completeness of the answer. Finally, the \textbf{Utility}, denoted as $A_u$, determines the usefulness of the answer.  The human score of an answer is a linear combination of the three metrics: 
\begin{equation}
    S^h = [ \alpha_r A_t + \beta_r A_c + \gamma_r A_u ] \: ; \: \alpha_r + \beta_r + \gamma_r = 1
\end{equation}

\item \textbf{User Constructs:} encompass parameters that signify a user's proficiency and ability in evaluating the generated text, showcasing the quality of their assessment and its influence on the overall human score (\ie calculate $\omega^h$). To do so, we define four metrics. First, \textbf{Duration}, denoted as $D$, measures the time interval in minutes between the last two evaluations. Second \textbf{Familiarity}, denoted as $F$, gauges the user's familiarity with the response context. Third, \textbf{LLM Trust}, denoted as $T$, assesses the user's belief in the expertise of the targeted LLM. Finally, \textbf{Uncertainty}, denoted as $U$, captures the user's degree of uncertainty regarding the evaluation. The weight of the human evaluation is given by:
\begin{equation}
  \omega^h = \mathcal{W}^h \mathcal{F}_D  
\end{equation}
Where,
\begin{equation*}
  \mathcal{W}^h =  [ \alpha_u F + \beta_u T + \gamma_u (1-U) ]  \: ; \: \alpha_u + \beta_u + \gamma_u = 1
\end{equation*}
and,
\begin{equation*}
   \mathcal{F}_D = tanh_{\lambda}(D) =  \frac{1-e^{-\lambda.D}}{1+e^{-\lambda.D}} 
\end{equation*}

We normalize $D$ using a hyperbolic tangent function $\mathcal{F}_D\in [0,1]$. $\mathcal{F}_D$ is implemented in a way that thwarts potential abuse. It reduces the impact of successive evaluations performed within a short period, thereby protecting the LLM's overall reputation and reinforcing the model's effectiveness. Furthermore, the positive correlation with the other metrics (\ie $F$, $T$, and $1-U$ ) leads to important considerations: first, ratings from users less familiar with the context carry less weight in updating the model's overall human reputation; second, ratings from users with minimal trust or with higher uncertainty have less impact on updates compared to those with lower uncertainty and higher trust in the overall expertise of LLMs.
\end{itemize}

\subsection{Overall Scores Update} \label{subsec:repupdate}
\quad In LLMChain, we employ three types of updates. The overall automatic reputation $R^{a}$ update occurs after each interaction to keep tracking the LLM behavior, while changes in $R^{h}$ and $R$ only occur if the interaction includes a human evaluation. These updates depend on the outcome of the automatic evaluation $S^a$, the human evaluation $S^h$, or the weighted evaluation $S^{\theta}$. We use $\theta$, a configurable weighting factor, to give more emphasis to the human evaluation when calculating $S^{\theta}$ and $\omega^{\theta}$, as follows:
\begin{equation}\label{eq:convex}
\left\{ \begin{array}{cl}
S^{\theta} = \theta S^h + (1-\theta)S^a 
\\ \omega^{\theta} = \theta \omega^h + (1-\theta)\omega^a
\end{array}\right.
\end{equation}
The updating formula $\mathcal{U}_{\psi,\xi}: (R_{i},~S_{calc},~ \omega) \longrightarrow R_{i+1}$ for the three scores $R^{h}$, $R^{a}$, and $R$ is thus defined as follows: \\
\\$\forall (R_{i}, S_{calc},\omega) \in \{(R^{a},S^a,\omega^a), (R^{h},S^h,\omega^h),(R, S^{\theta},\omega^{\theta})\},$
\begin{equation}\label{eq:globalupdate}
R_{i+1} =  \left\{ \begin{array}{cl}
(1- \psi \omega) R_{i} + \psi \omega S_{calc} \: ; & S_{calc} \geq \overline{R_i} 
\\(1- \xi \omega) R_{i} + \xi \omega S_{calc} \: ; & S_{calc} <  \overline{R_i}
\end{array}\right.
\end{equation}
where $R_{i}$ and $\overline{R_i}$ are the current reputations and trust thresholds (\ie before the interaction $i$), respectively. We define the threshold $\overline{R_i}$ as the average of LLM reputations. 

\quad By employing two distinct formulas in (Eq. \ref{eq:globalupdate}) for the update process using a trust threshold $\overline{R_i}$, we separate expected good behavior from unexpected bad behavior (no/bad response, hallucination, harmful content, etc.~\cite{infocomcrowdrep,repsrv}). Consequently, we can put more weight (\ie $\xi > \psi$) on the newly calculated score $S_{calc}$ in the case of an incorrect response. Moreover, the integration of the weighting function $\omega$ into both equations establishes a direct relationship between the quality of the evaluation and its impact on the update of the overall reputation. For instance, for a $R^h$ update, the greater the user's familiarity, certainty, and trust in the LLM expertise, the more significant their evaluation's impact becomes. Moreover, the use of $D$ allows the system to mitigate consecutive inaccurate ratings that may be intended to enhance or damage LLM's reputation. We note that this metric is reset at regular intervals (\eg every 24 hours), preventing users who abstain from evaluations for a long time from exploiting the model.
\begin{figure}[t]
   \centering
    \subfloat[Reputation growth following successive accurate answers]{
      \includegraphics[scale=0.3]{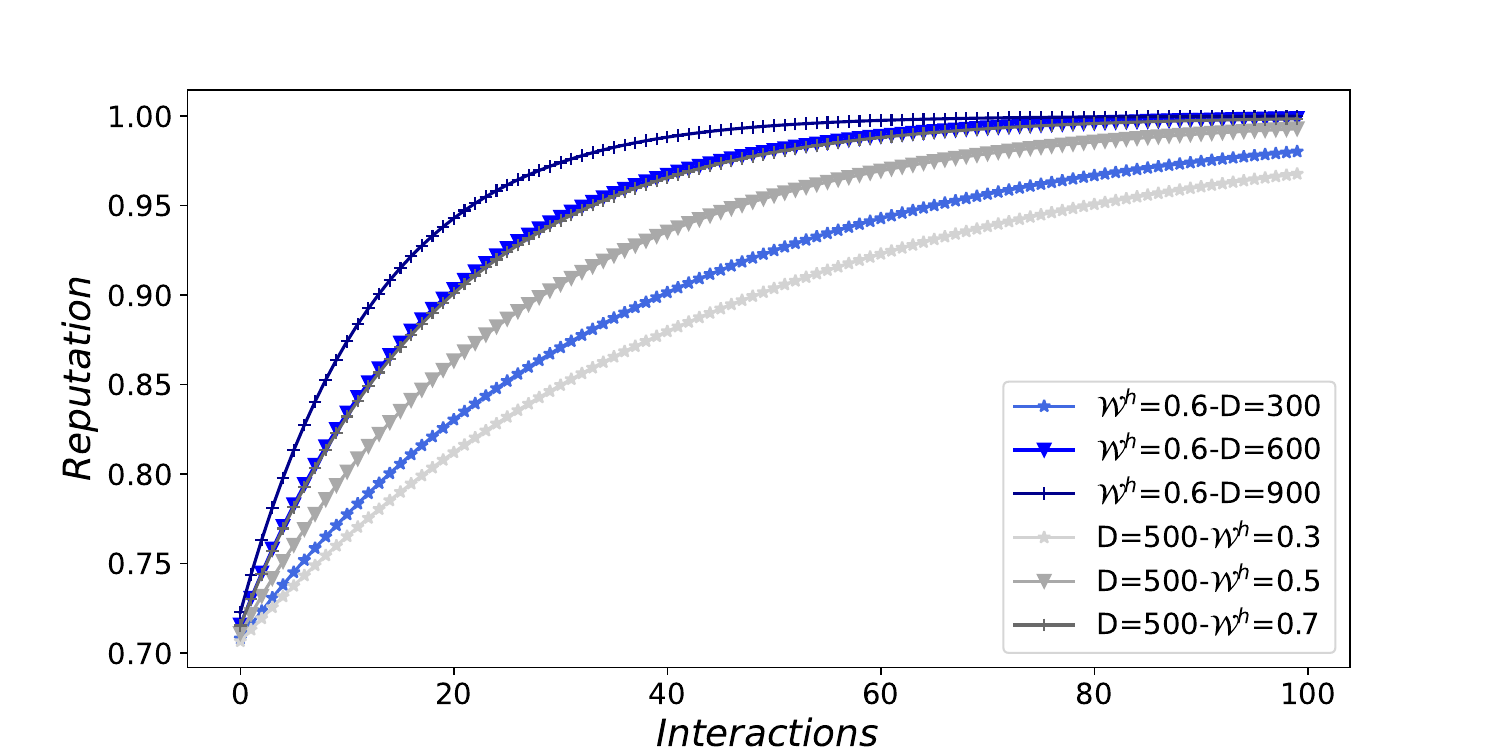}
       \label{fig2a}
    }
    \hfill
   \subfloat[Reputation changes after successive incorrect answers]{
       \includegraphics[scale=0.3]{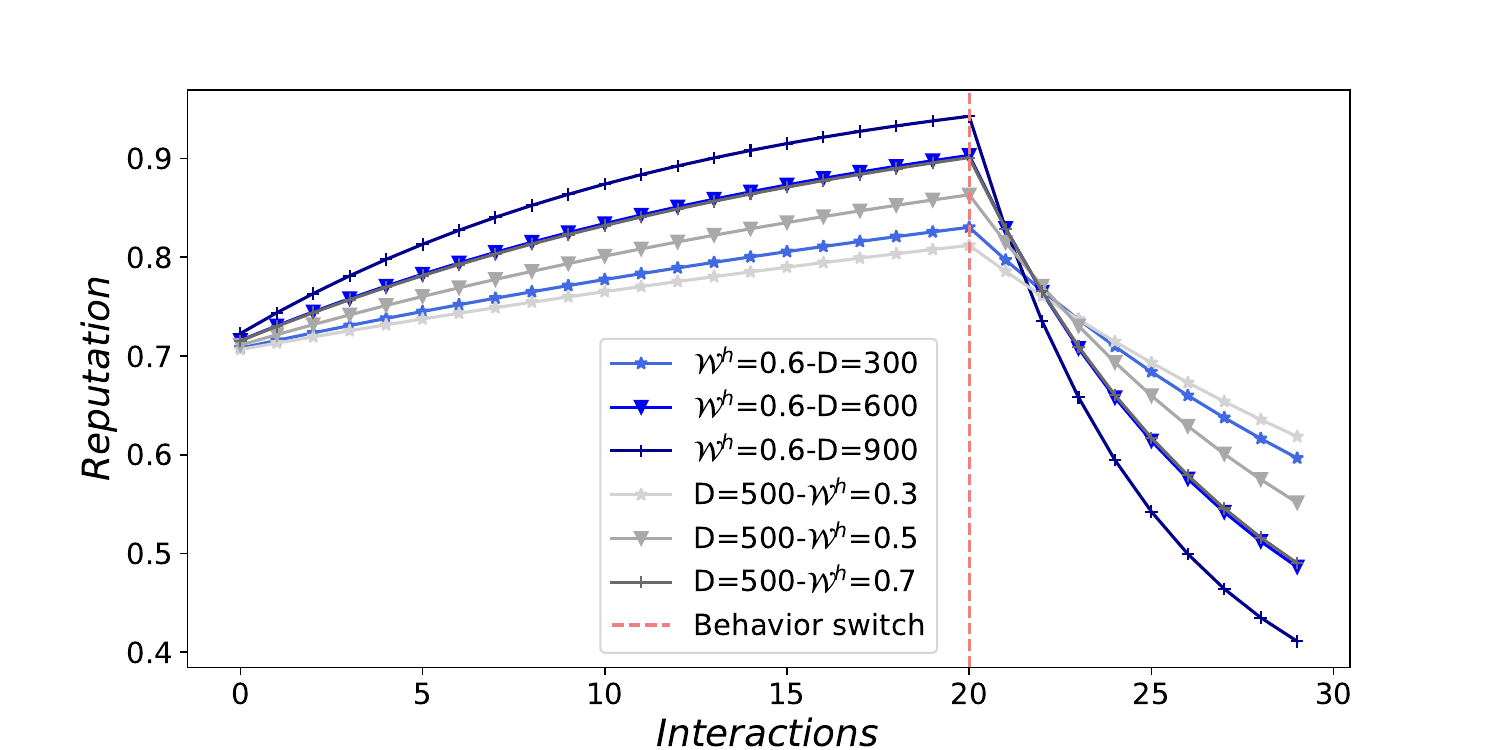}
       \label{fig2b}
    }
    \caption{The Effectiveness of LLMChain's Reputation model under different $\mathcal{W}^h$ and $D$.}
    \label{fig2}
\end{figure}

\quad Fig.~\ref{fig2} demonstrates the impacts of $D$ and $\mathcal{W}^h$ on the overall reputation updates. It shows the shifts in reputation between a skilled model consistently providing accurate responses and a less competent one that produces consecutive incorrect answers after delivering multiple correct ones. Both positive and negative updates have a direct correlation with $D$ and $\mathcal{W}^h$. This suggests that the longer the time interval between the last two evaluations, the more significant impact the user's latest evaluation has. Likewise, increased levels of familiarity, trust, and certainty contribute to a more substantial impact.

\section{Experiments} \label{sec:proofOfConcept} 


\subsection{Experimental Setup} 

\subsubsection{Environment} We conducted the experimental tests on two separate clusters: a GPU cluster for hosting the LLM part of the system and a CPU cluster dedicated to running the blockchain network. The first cluster comprises two servers, one featuring an NVIDIA RTX A6000 GPU card and the other equipped with an NVIDIA GeForce RTX 2080 Ti card. Meanwhile, the second cluster consists of two HPE ProLiant XL225n Gen10 Plus servers specifically allocated for experimenting with blockchain solutions. Each server in this cluster is powered by two AMD EPYC 7713 64-Core processors and 2x256 GB RAM.

\subsubsection{Datasets} We evaluate LLMChain on three datasets:
\begin{itemize}
    \item \textit{MTBench}\footnote{\href{https://huggingface.co/spaces/lmsys/mt-bench}{https://huggingface.co/spaces/lmsys/mt-bench}} is a recent dataset extensively utilized in evaluating LLMs\cite{llmjudge}. MT-Bench consists of 3.3K expert-level pairwise human preferences for answers generated by six models (``Llama-13B'', ``Alpaca-13B'', ``Vicuna-13B'', ``GPT-3.5'', ``Claud-v1'', and ``GPT-4'') across 80 questions.
    \item \textit{GooAQ}\footnote{\href{https://huggingface.co/datasets/gooaq}{https://huggingface.co/datasets/gooaq}} is a large-scale dataset with a variety of answer types. This dataset comprises more than 5M questions and 3M answers sourced from Google \cite{gooaq}.
    \item \textit{LLMGooAQ.}\footnote{\href{https://github.com/mohaminemed/LLMGooAQ/}{https://github.com/mohaminemed/LLMGooAQ/}} We prepare this comprehensive database, covering 100k questions and answers in 20 different fields/contexts. We randomly sample 100K tuples from the GooAQ dataset and perform inference using seven LLMs (``Alpaca-13b'', ``Llama-2-13b'', ``Chatglm-6b'', ``Fastchat-t5-3b'', ``Koala-13b'', ``Vicuna-7b'', ``Vicuna-13b'').
\end{itemize}

\subsubsection{Automatic Metrics} \label{subsec:autometrics} To pinpoint the optimal technique for our context, we conduct rigorous benchmarks among various embedding-based metrics that achieved SoTA performance.
\begin{itemize}
\item \textbf{BERTScore} \cite{bert} is an automatic evaluation metric for text generation. It evaluates the similarity between tokens in a candidate sentence and those in a reference sentence. Unlike N-Gram methods relying on exact matches like BLEU Score\cite{bleu} and ROUGE Score\cite{rouge}, BERTscore relies on contextual embeddings to gauge token similarity. The approach employs cosine similarity to measure the likeness between a reference token $x_i$ and a candidate token $\hat{x}_i$. The total score involves comparing each token in $x$ with tokens in  $\hat{x}$ to calculate recall, and each token in  $\hat{x}_i$ with tokens in $x$ to determine precision. To maximize the similarity score, a greedy matching technique is used, wherein each token is paired with the most similar token from the other sentence. Precision and recall are combined to derive an F1 score. 



\item \textbf{BARTScore} \cite{bart} is an automated evaluation method that frames the evaluation of generated text as a text generation problem, utilizing pre-trained sequence-to-sequence models. The fundamental concept revolves around the notion that models trained to convert generated text into or from a reference output or the source text will yield higher scores for superior generated text. This concept is implemented using BART, a pre-trained model based on an encoder-decoder architecture. The metric BARTScore offers various adaptable variants that can be applied in an unsupervised manner to evaluate text from multiple perspectives, such as informativeness, fluency, or factuality. 




\item \textbf{DISCOScore} \cite{disco} is a parametrized discourse metric, which uses BERT to model discourse coherence from different perspectives, through the lens of readers’ focus, driven by Centering theory. DISCOScore offers two variations: FocusDiff and SentGraph, differing in their treatment of focus. This approach models the frequency and semantic relevance of focus and then compares the disparities between the hypothesis and the reference. It utilizes two adjacency matrices to represent coherence based on focus. In FocusDiff, the matrix represents relationships between foci and tokens, indicating focus frequency. Meanwhile, in SentGraph, the matrix showcases the interdependence between sentences based on shared foci and sentence proximity.  
\end{itemize}

\begin{figure*}[t]
    \centering
    \subfloat[DSFocus]{
        \includegraphics[scale=0.27]{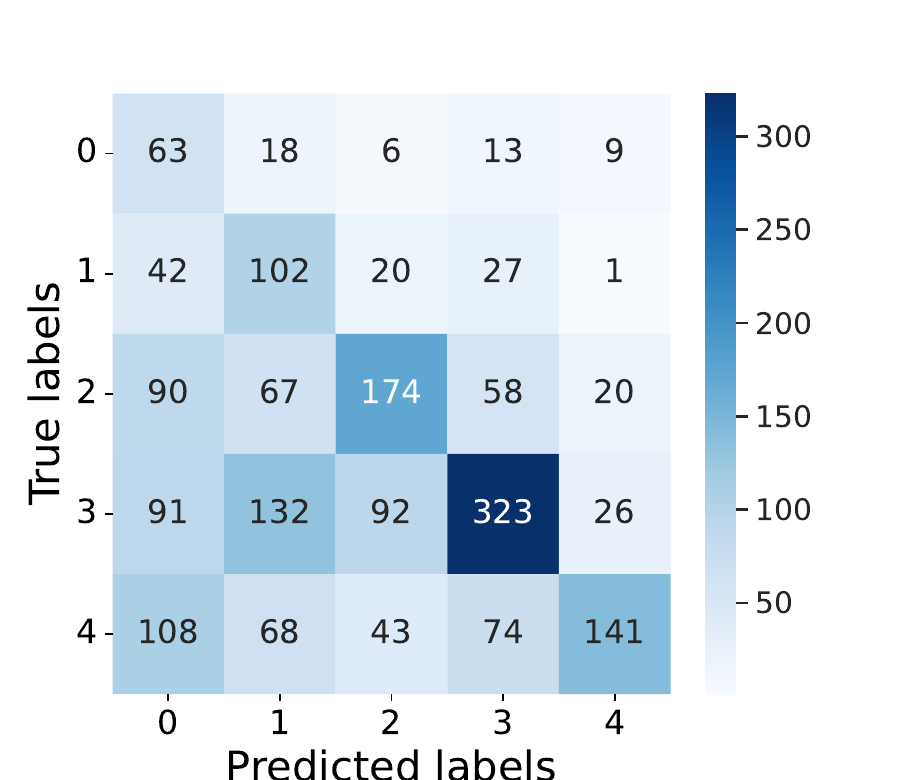}
        \label{fig3a}
    }
    \hfill
    \subfloat[DSSent]{
        \includegraphics[scale=0.27]{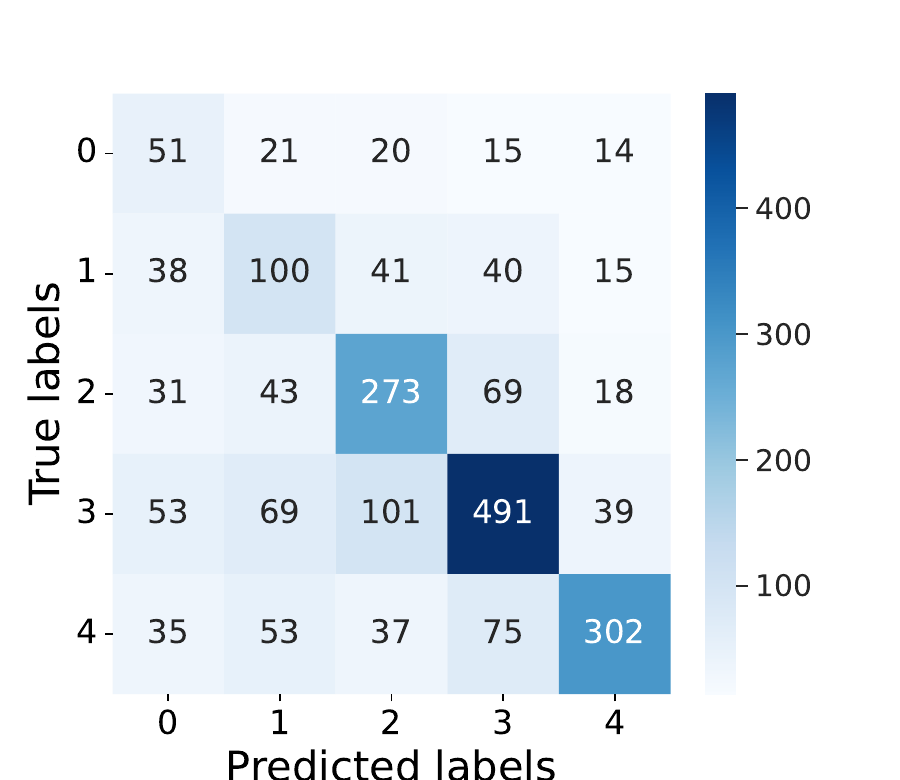}
        \label{fig3b}
    }
   \hfill
    \subfloat[BERTScore]{
        \includegraphics[scale=0.27]{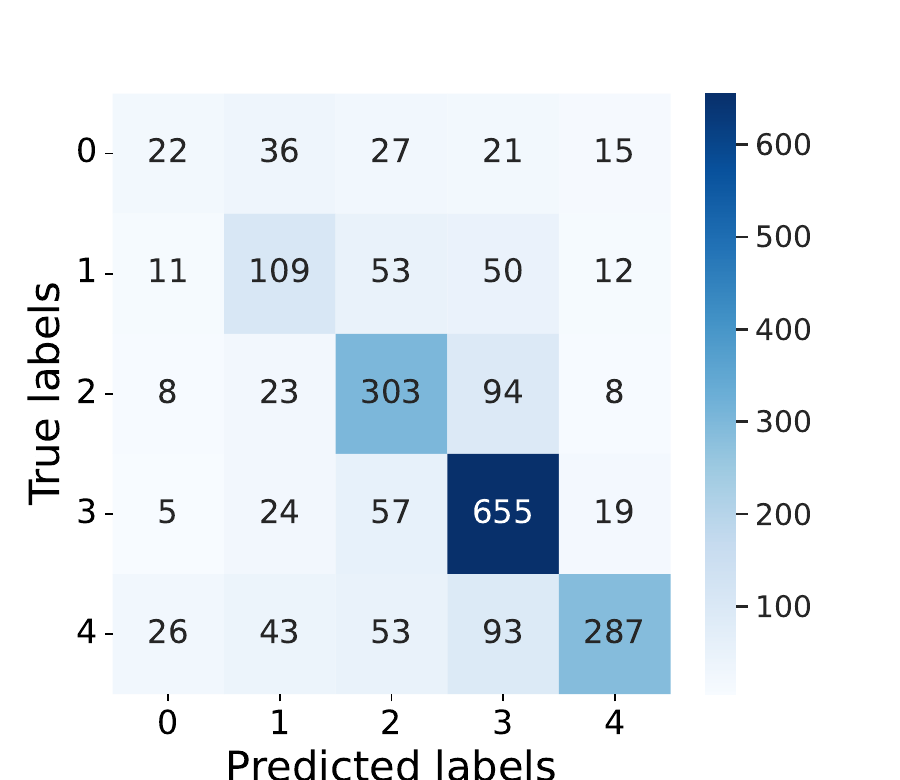}
        \label{fig3c}
    }
     \hfill
    \subfloat[BARTScore]{
        \includegraphics[scale=0.27]{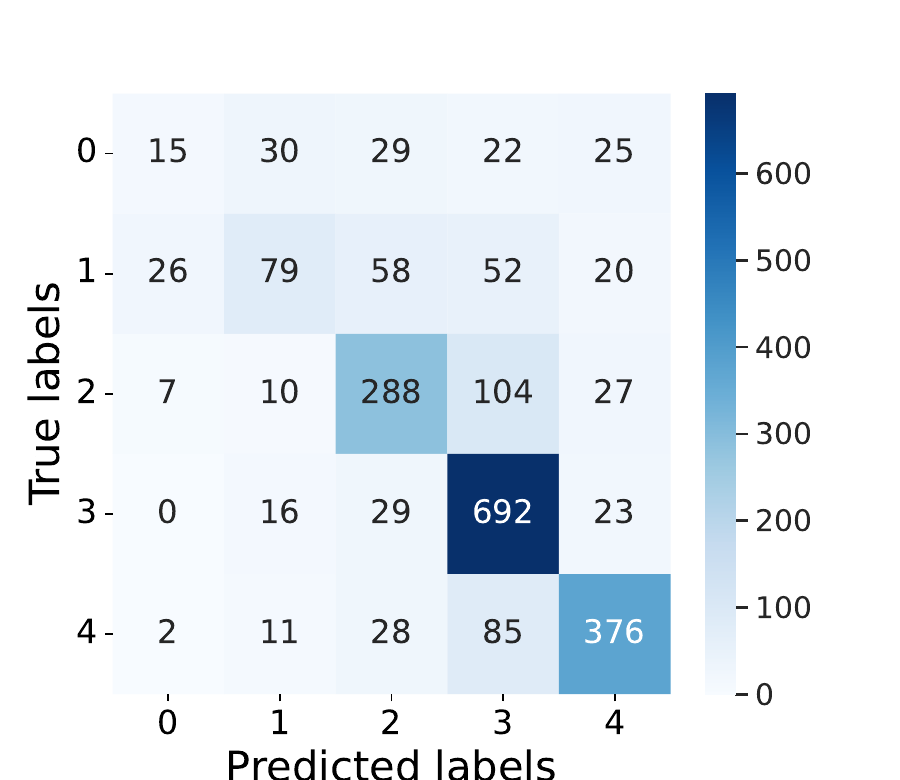}
        \label{fig3d}
    }
    \caption{\centering -. Labels are denoted as:
    \{``0:Llama-13B'', ``1:Alpaca-13B'', ``2:Vicuna-13B'', ``3:GPT-3.5'', ``4:Claud-v1''\}}.
    \label{fig3}
\end{figure*}

     

\begin{table}[t]
\centering
\caption{Hyperparameter's Configuration.}
\resizebox{.9\columnwidth}{!}{%
\begin{tabular}{|l|c|c|c|}
    \hline
     Parameter & Value & Parameter & Value \\
    \hline
     $\psi$ &  $1/3$ &  $\alpha_r=\beta_r=\gamma_r$ &  $1/3$  \\
     $\xi$ &  $2/3$  &  $\alpha_u=\beta_u=\gamma_u$ &  $1/3$ \\
     $\lambda$ &  $10^{-3}$ & $\theta$ &  $2/3$  \\ 
    \hline
\end{tabular}%
}
\label{tab1}
\end{table}


\subsection{Reputation Model Effectiveness} \label{sec:repeval}

\quad In the following, we first perform an experimental comparison of the automatic metrics described in Sec.~\ref{subsec:autometrics}. Next, we perform two additional experiments aiming to evaluate the efficiency of both the automatic and human models. The values of the configurable parameters used in these experiments are summarized in Table.~\ref{tab1}.

\begin{enumerate}
\item \textbf{Metrics Benchmark.} Determining the most fitting metric for evaluating LLM-generated answers analytically is not straightforward. That is why we embarked on a benchmark experiment to pinpoint the best technique. This experiment aims to assess the metrics commonly used in automatically evaluating NLP tasks. Our goal is to identify the one that best aligns with human judgments. To achieve this, we conduct an experiment that involves computing automatic scores on MTBench answers. These scores automatically determine the winner between two different LLMs for each question. Fig.~\ref{fig3} demonstrates the correlation between human-selected winners (true) and automatic winners (predicted). The matrices show nearly diagonal patterns, indicating good correlations, yet variations in accuracy exist. For instance, the DISCOScore DSSent variant boasts an accuracy of 59\%, surpassing that of the DSFocus variant (44\%). BARTScore, on the other hand, demonstrates superior accuracy, with 71\% of predicted winners matching actual human winners, compared with 67\% for BERTScore. Table \ref{tab2} illustrates Kendall's Tau correlation of these four metrics. We can see that BARTScore can significantly outperform all other techniques by offering a superior correlation of 80\% with human judgments. Based on these results, we decided to use \textbf{BARTScore} in the following experiments.

\begin{table}[t]
\centering
\caption{Metrics Performance on the MTBench dataset.}
\resizebox{.7\columnwidth}{!}{%
\begin{tabular}{|l|c|c|}
    \hline
     Metric & Accuracy &  Kendall's Correlation  \\
    \hline
     DSFocus &  0.44414   &  -0.60  \\ 
     DSSent &  0.59540   & 0.60 \\     
     BertScore &  0.66991    & 0.60 \\
     BartScore & \textbf{0.70594}   & \textbf{0.80} \\
    \hline
\end{tabular}%
}
\label{tab2}
\end{table}

\item \textbf{Automatic Evaluation.} To adequately evaluate the automatic model, we use BARTScore to conduct a pairwise comparison between the seven LLMs in LLMGooAQ using GooAQ's answers as benchmarks. Subsequently, we calculate the win rates for each LLM per context. The experimental results, showcased in Fig.~\ref{fig7}, highlight ``Vicuna-13b'' as the best model outperforming others in nearly 90\% of the contexts. Furthermore, the resulting models' overall win rates align with previous human-based evaluation~\cite{llmjudge}, affirming that the BARTScore metric correlates strongly with human judgments.

\quad Now, to assess the efficacy of leveraging the best models' answers within specific contexts, we conduct a subsequent test using the answers from ``Vicuna-13b'' as references. Fig. \ref{fig8} presents the confusion matrix comparing the winners (true) computed using GooAQ answers with those (predicted) computed using ``Vicuna-13b'' answers. The results are compelling, revealing robust accuracy (70\%) between the two cases. It is essential to note that, according to current benchmarks \cite{llmjudge,trustllm} and leaderboards (ChatBotArena\footnote{\href{https://huggingface.co/spaces/lmsys/chatbot-arena-leaderboard} {https://huggingface.co/spaces/lmsys/chatbot-arena-leaderboard}}, TrustLLM\footnote{\href{https://trustllmbenchmark.github.io/TrustLLM-Website/leaderboard.html}{https://trustllmbenchmark.github.io/TrustLLM-Website/leaderboard.html}}), ``Vicuna-13b'' is a well-ranked open source model, but it is not the best. Despite this, the results obtained using it as a reference model are convincing.  

\begin{figure}[t]
\centering
\includegraphics[width=0.53\textwidth]{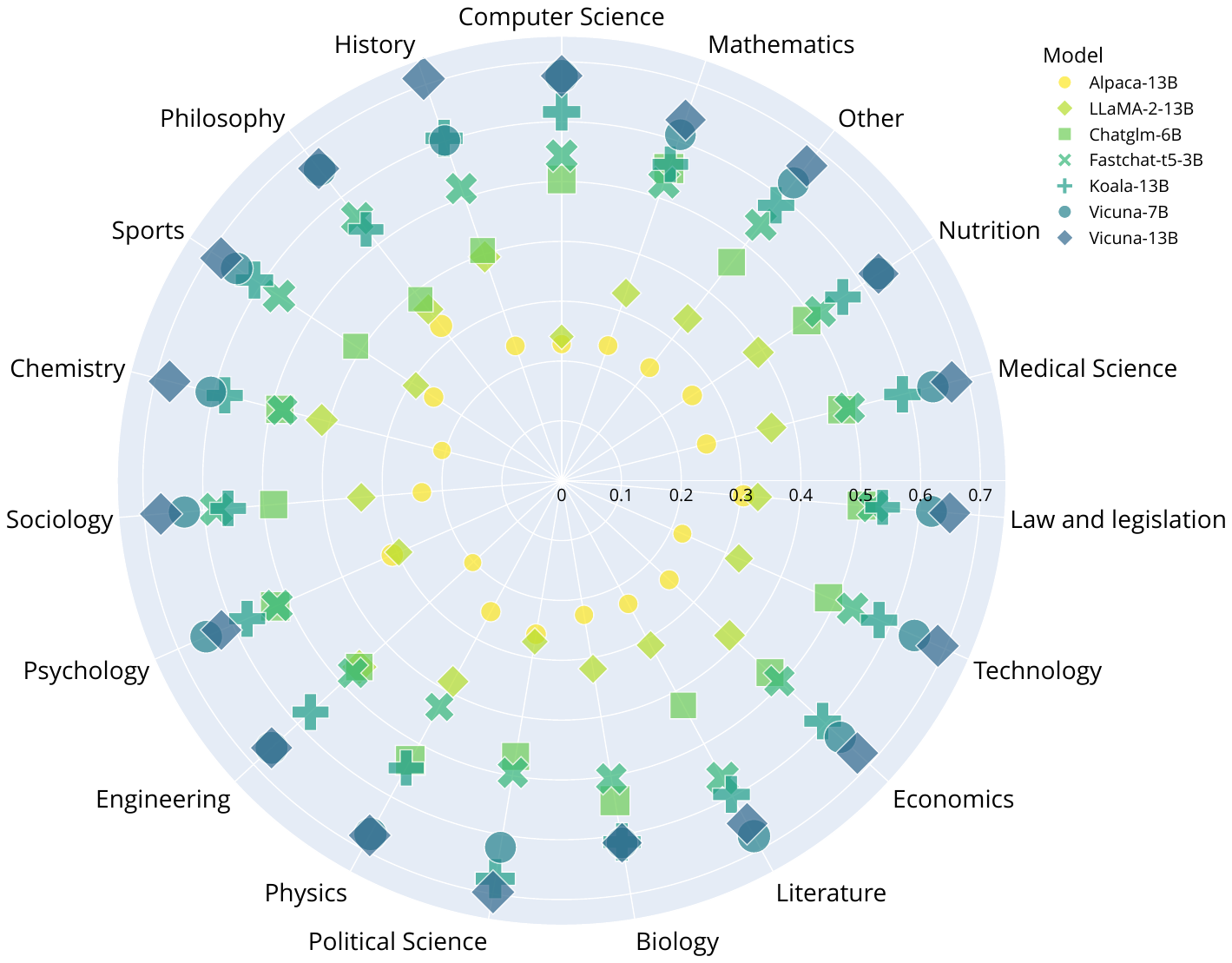}
\caption{BARTScore-based Contextual Win-Rates on LLMGooAQ.}
\label{fig7}
\end{figure}

\item \textbf{Reputation Evaluation.} The third experiment involves employing the proposed models and monitoring changes in reputations in a real scenario. We use our prepared dataset with automatic scores computed using BARTScore to do this. Given the high cost of obtaining human judgments, we employ GPT-4 as an expert for human evaluation. GPT-4 is recognized as the leading model in current benchmarks \cite{llmjudge,trustllm,coregpt}. In this experiment, GPT-4 is used to play the role of a human expert, responding to a questionnaire that enables the calculation of metrics (\ie $F$, $T$, $U$, $A_t$, $A_c$, and $A_u$) used in the human model. Fig. \ref{fig9} illustrates the variations in $R^{a}$, $R^{h}$, and $R$ for the seven LLMs in our dataset. Despite the disparities between the $R^{a}$ and $R^{h}$ scores, a consistent pattern emerges, with scores for good models such as ``Koala-13b'', ``Vicuna-7b'', and ``Vicuna-13b'' steadily increasing, while scores for less effective models such as ``Alpaca-13b'' and ``Llama-2-13b'' continually decrease. Moreover, with an increasing number of evaluations, the distinctions between closely ranked models become more pronounced. This demonstrates the effectiveness of our models, showcasing their ability to discern even subtle differences between close LLMs like ``Chatglm-6b'' and ``Fastchat-t5-3b''.     
\end{enumerate}

\begin{figure}[t]
\centering
\includegraphics[width=0.4 \textwidth]{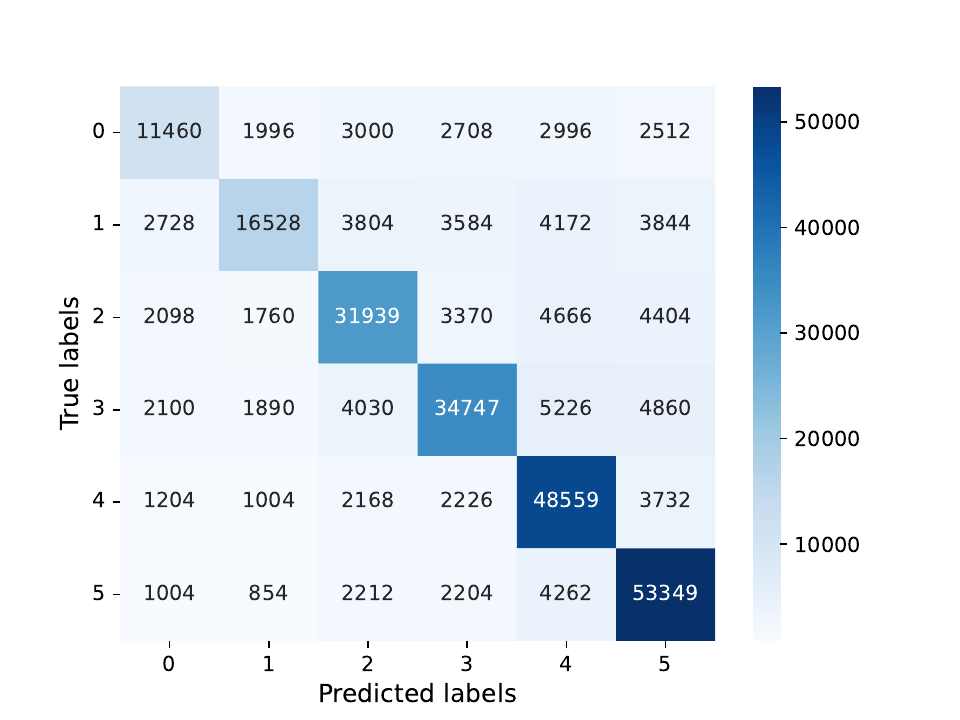}
\caption{Ground-Truth Answers vs Vicuna-13B Answers as References for BARTScore-based Pairwise-comparison on the LLMGooAQ dataset. Labels are denoted as:  \{0: ``Alpaca-13b'', 1: ``Llama-2-13b'', 2: ``Chatglm-6b'', 3: ``Fastchat-t5-3b'', 4: ``Koala-13b'', 5: ``Vicuna-7b'', 6: ``Vicuna-13b''\}.}
\label{fig8}
\end{figure}

\subsection{Blockchain Performance}
\subsubsection{Business Model} Having evaluated all its components in the previous subsection, we now implement the proposed blockchain-driven framework. This one is deployed on a blockchain network powered by Hyperledger Besu\footnote{\href{https://besu.hyperledger.org}{https://besu.hyperledger.org}}, an open-source Ethereum client. Our evaluation approach includes:
\begin{itemize}
\item Participants: Users with different expertise and Admins of the organization or the consortium operating the system. 
\item Assets: A data structure that represents the model on-chain.
\item Smart Contracts: Three types of smart contracts are used to develop the business model: Identity Smart Contract (ISC), Access Control Smart Contract (ACSC), and  Reputation Smart Contract (RSC). ISC implements the registration process, ACSC employs a role-based access control to manage the permissions when calling RSC functions, \eg only Oracles can trigger the $autoEval$ function.  The RSC implements four main functions, $addModel$, $autoEval$, $humEval$, and $updateReputation$.
\end{itemize}
\quad We develop the smart contracts of LLMChain using the $\mathsf{Solidity}$ programming language\footnote{\href{https://docs.soliditylang.org}{https://docs.soliditylang.org}} and establish a local network consisting of sixteen validators using Hyperledger Besu with Proof of Authority (PoA) as consensus protocol. We lastly use  $\mathsf{Web3js}$ library\footnote{\href{https://web3js.readthedocs.io/}{https://web3js.readthedocs.io}} for developing the client side and deploying the system's smart contracts. 


\begin{figure*}[t]
    \subfloat[BARTScore-based Automatic scores]{
        \includegraphics[scale=0.23]{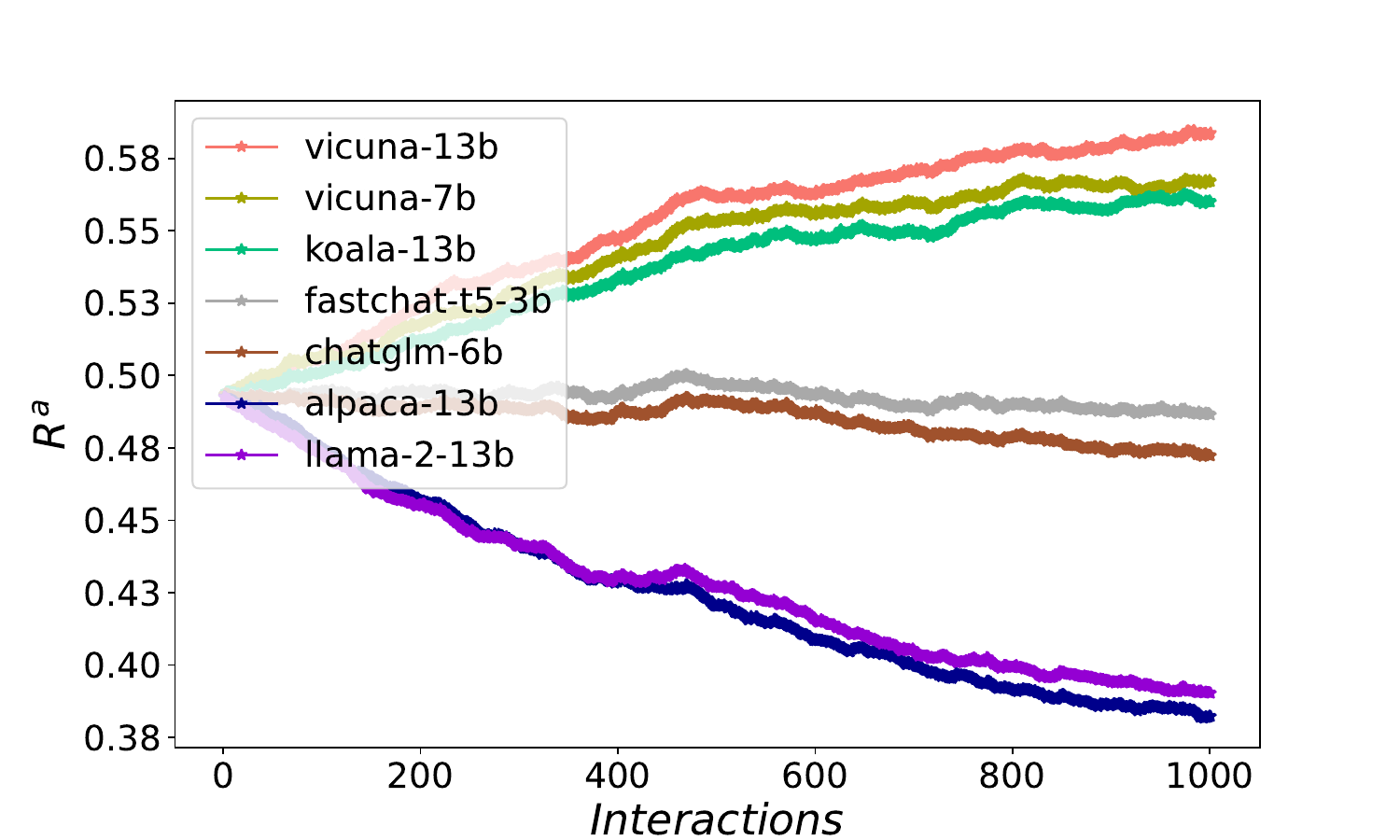}
        \label{fig9a}
    }
    \subfloat[GPT4-generated Human scores]{
        \includegraphics[scale=0.23]{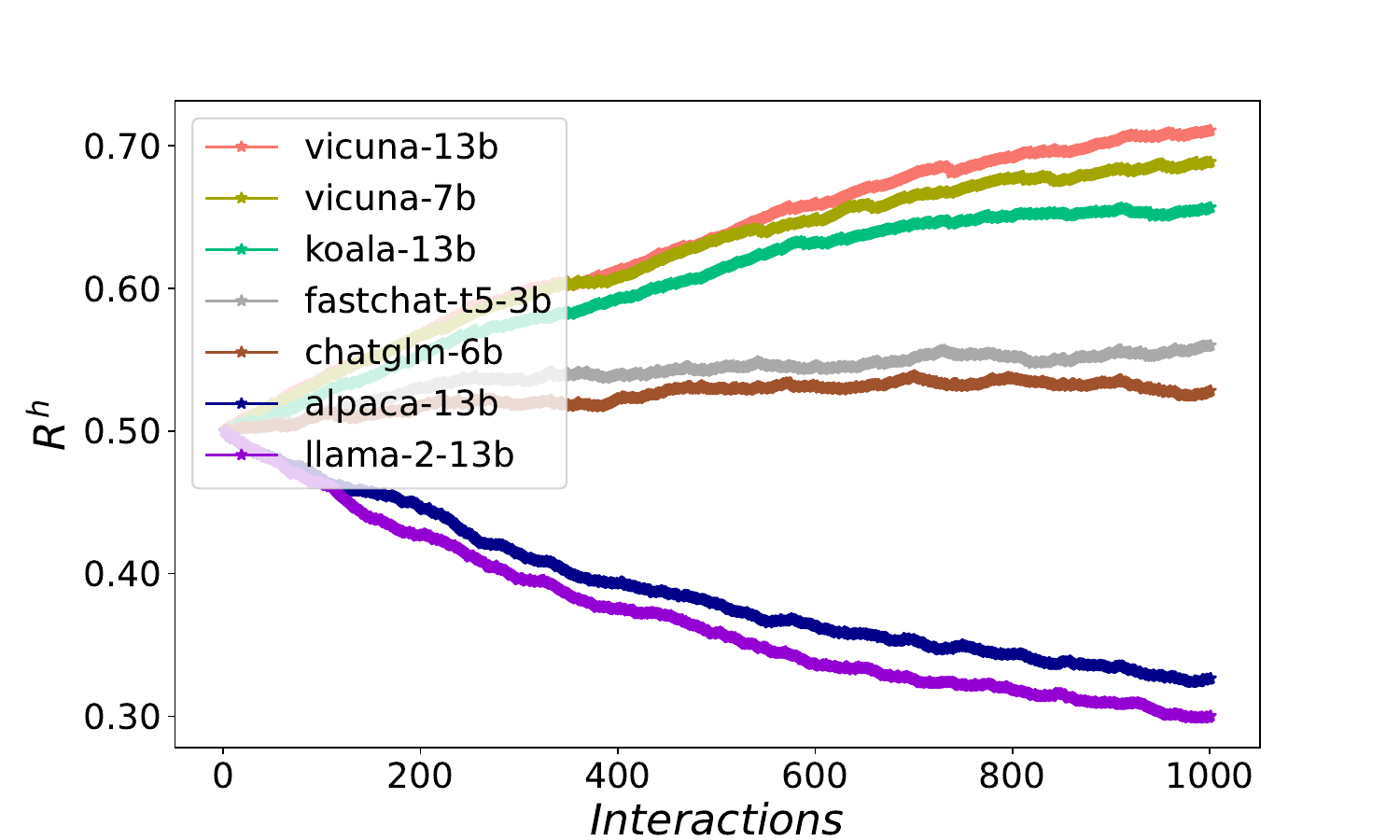}
        \label{fig9b}
    }
    \subfloat[Weighted Reputation scores]{
        \includegraphics[scale=0.23]{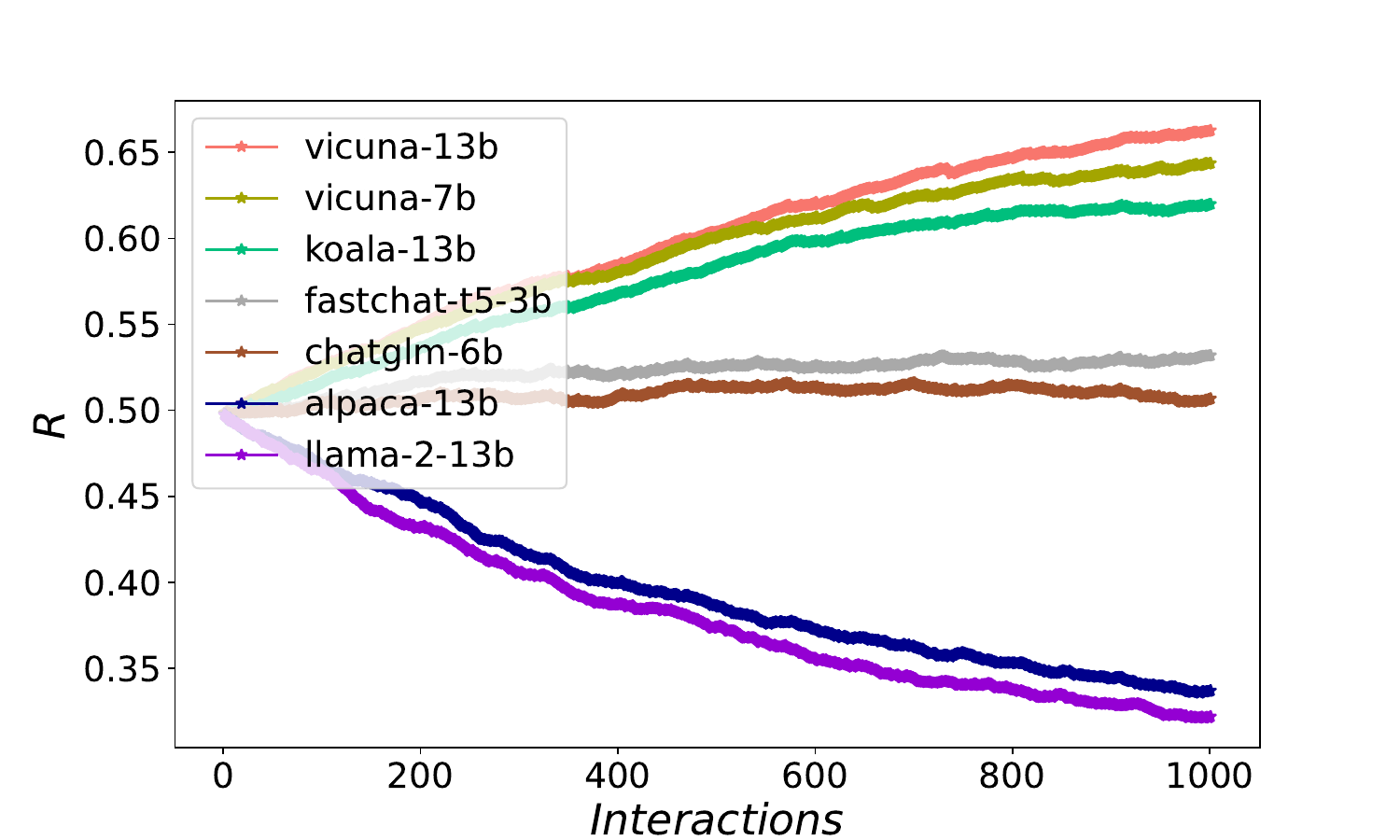}
        \label{fig9c}
    }
  
    \caption{\centering Changes in $R^{a}$, $R^{h}$, and $R$ of seven LLMs using LLMGooAQ.}
    \label{fig9}
\end{figure*}

\begin{figure}[t]
    \centering
    \subfloat[Throughput under different send rates]{
        \includegraphics[scale=0.3]{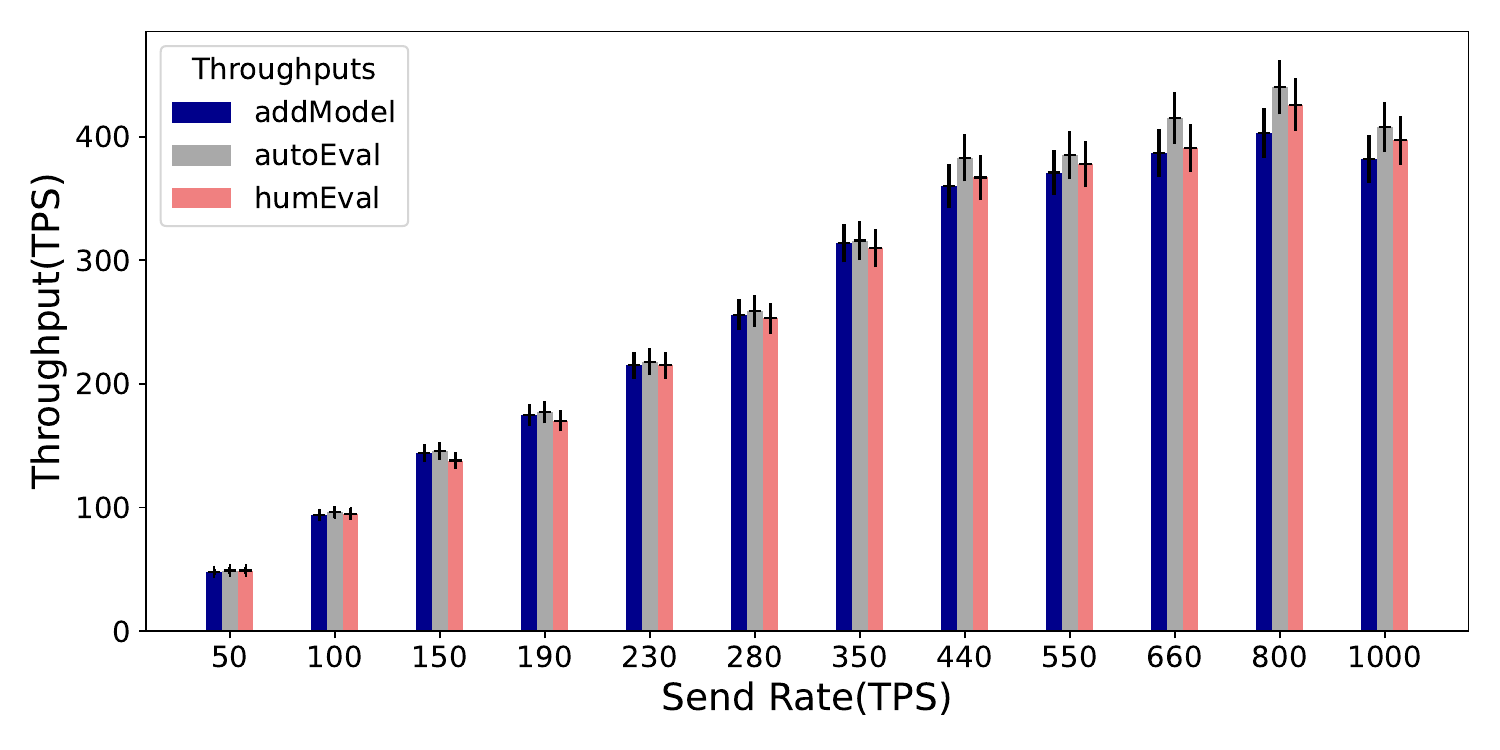}
        \label{fig10a}
    }
    \hfill
    \subfloat[Latency under different send rates]{
        \includegraphics[scale=0.3]{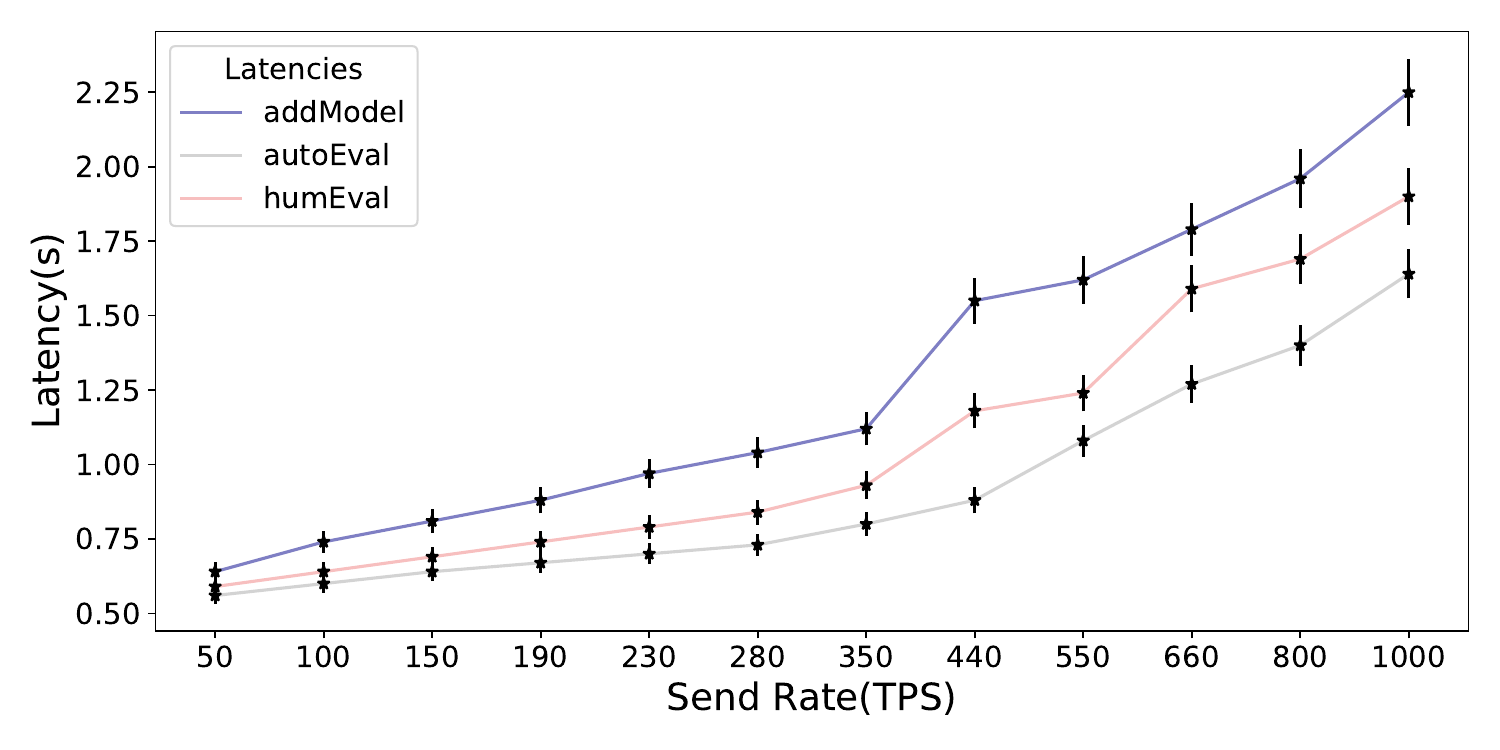}
        \label{fig10b}
    }
  
    \caption{Throughput and Latency of LLMChain.}
    \label{fig10}
\end{figure}

\subsubsection{Performance Evaluation} To conduct tests, we utilize Hyperledger Caliper\footnote{\href{https://github.com/hyperledger/caliper-benchmarks}{https://github.com/hyperledger/caliper-benchmarks}}, a benchmarking tool for blockchains. The experiments involve changing the transaction sending rate (ranging from 50 to 1000 TPS) using a consistent network configuration for the main operations performed within LLMChain. As a result, two metrics are measured:
\begin{itemize}[leftmargin=0cm,align=left]
\item \textbf{Throughput:} is the number of successful transactions per second (TPS).
\item \textbf{Latency:} refers to the time difference in seconds between the submission and completion of a transaction.
\end{itemize}

\quad The throughput and latency values for each function under different sending rates are illustrated in Fig.~\ref{fig10}. At the beginning, the pattern is evident: throughput and latency increase as the transaction send rate increases. With lower sending rates ($<$350 TPS), there is no significant difference in throughput between the three defined transaction types. However, nearing system capacity, distinctions emerge. The lightest function, $autoEval$, achieves a peak throughput of $440$ TPS, surpassing $humEval$ at $426$ TPS, and the heaviest function, $addModel$, managing $403$ TPS, primarily due to the initialization and storage of model information on-chain. This also explains the comparatively higher latency of $addModel$ compared with the other two functions. Nevertheless, leveraging storage scaling via IPFS, LLMChain achieves an average throughput close to $420$ TPS, comfortably meeting the specific demands of our use case. On top of that, since LLMChain operates on an EVM-based state machine, all the scaling techniques of Ethereum-based blockchains, such as Sharding and zkRollups can be applied to further enhance its performance for large-scale deployment if needed. 

\section{Limitations and Future research directions}
\quad To the best of our knowledge, we are the first to design and develop a reputation model for evaluating LLMs within a decentralized framework. While our experiments prove the effectiveness and scalability of LLMChain, we believe that this work promotes future research on decentralized and transparent language model evaluation. However, LLMChain presents some limitations regarding both human and automatic evaluations. Firstly, human evaluation depends mainly on users' willingness to provide authentic feedback. Further assurance and incentive measures can be added to the framework to improve the reliability of human evaluation. Secondly, automatic evaluation relies on the availability of reference models. This approach has proved effective. However, it has two important shortcomings: i) Its accuracy depends on the performance of available reference models, ii) and even if the $k$ responses can help the user to provide a better human evaluation, this approach generates off-chain communication and computational overheads. 
\normalsize
\section{Conclusion}
\label{sec:conclusion}
\quad In this paper, we propose LLMChain, a novel blockchain-powered framework, specifically designed to share and evaluate LLMs efficiently and transparently. LLMChain addresses trust concerns associated with flawed behaviors like hallucinations and unreliable reasoning of LLMs by employing a context-driven reputation system. Our efforts involve crafting and implementing a reputation model that evaluates user satisfaction and trust in each interaction involving an LLM. This model amalgamates human feedback with automatic evaluation to assign contextual reputation scores that accurately mirror LLM behavior. Consequently, the system aids users and entities in pinpointing the most credible LLM for their requirements while offering LLM providers valuable insights to refine and enhance their models. This research marks the first initiative to introduce a distributed framework dedicated to LLMs evaluation. Through extensive experiments and benchmarks, we demonstrate the effectiveness of both human and automatic evaluations in LLMChain. Moreover, the tests conducted on the deployed blockchain affirm LLMChain's efficiency and scalability, validating its practical applicability in real-world scenarios. 
Finally, LLMGooAQ, a large dataset of over 100K questions and answers generated using seven LLMs, was prepared and released to the community to advance research in this area further.

\section*{acknowledgement}
{\scriptsize
This work was supported by the 5G-INSIGHT bilateral ANR-FNR project 
, the Nouvelle-Aquitaine Region - B4IoT project, the French government in the framework of the France Relance program, and the ITSOFT company under grant number AD 22-252.
}


\bibliographystyle{IEEEtran}
\bibliography{ref}

\end{document}